\documentclass[fleqn,twoside]{article}

\usepackage{amsmath,amssymb}

\usepackage{espcrc2}

\usepackage[dvips]{graphicx}

\allowdisplaybreaks

\title{The scaling equation of state of the three-dimensional $O(N)$ universality class: $N\geq 4$.}

\author{Agostino Butti\address{Dipartimento di Fisica dell'Universit\`a di Milano-Bicocca and INFN},
  Francesco Parisen Toldin\address{Scuola Normale Superiore and INFN}
  \thanks{\mbox{Speaker at the conference } \mbox{(Email: {\tt f.parisentoldin@sns.it})}},
  Andrea Pelissetto\address{Dipartimento di Fisica dell'Universit\`a di Roma ``La Sapienza'' and INFN},
  Ettore Vicari\address{Dipartimento di Fisica dell'Universit\`a di Pisa and INFN}
}

\begin{document}

\begin{abstract}
We determine the critical equation of state of the three-dimensional $O(N)$ universality class, for $N=4$, $5$, $6$, $32$, $64$. The $N=4$ is relevant for the chiral phase transition in QCD with two flavors, the $N=5$ model is relevant for the $SO(5)$ theory of high-$T_c$ superconductivity, while the $N=6$ model is relevant for the chiral phase transition in two-color QCD with two flavors. We first consider the small-field expansion of
the effective potential (Helmholtz free energy).
Then, we apply a systematic approximation scheme based
on polynomial parametric representations
that are valid in the whole critical regime,
satisfy the correct analytic properties (Griffiths' analyticity),
take into account the Goldstone singularities at the coexistence curve,
and match the small-field expansion of the effective potential.
From the approximate representations of the  equation of state,
we obtain estimates of universal amplitude ratios. We also compare our approximate solutions with those obtained in the large-$N$ expansion, up to order $1/N$, finding good agreement for $N\gtrsim 32$.
\end{abstract}

\maketitle

\section{Introduction}

In the theory of critical phenomena, continuous phase transitions
can be classified into universality
classes determined only by a few properties characterizing the system, such
as the space dimensionality,  the range of interaction,
the number of components of the order parameter,
the symmetry and the symmetry-breaking pattern \cite{ZJbook}.
Renormalization-group theory predicts that
critical exponents, universal amplitude ratios and scaling functions are
the same for all systems belonging to a given universality class.
The $O(N)$ universality classes are among the most important ones. 
They are characterized by an $N$-component order parameter and a symmetry
group $O(N)$ which is spontaneously broken to a subgroup $O(N-1)$ in
the low-temperature phase.
For a recent review on this subject, see ref.~\cite{review}.

Here we study the $O(N)$ three-dimensional universality class, with $N \geq 4$.
The three-dimensional $O(4)$ model is relevant for the
finite-temperature behavior of QCD with two light-quark
flavors \cite{PW-84}.
The $3$-$D$ $O(5)$ model is relevant for
the so-called $SO(5)$ theory of high-$T_c$ superconductivity \cite{SO5}.
According to universality arguments, the $N=6$ model should describe the chiral phase transition in two-color QCD with two flavors \cite{Smilga:1994tb}.

The equation of state is a relation between the magnetization
$\vec{M}$, the reduced temperature \mbox{$t \equiv (T-T_c)/T_c$} and the
external magnetic field $\vec{H}$. Near the critical point
it has the scaling \mbox{form \cite{ZJbook}}
\begin{equation}
\label{scaling}
\vec{H} = (B^c)^{-\delta} \vec{M} M^{\delta - 1} f(x),
\end{equation}
where $M\equiv |\vec{M}|$, $x \equiv B^{1/\beta}\, t M^{-1/\beta}$ is
a scaling variable, $f(x)$ is a universal scaling function,
fixed by the normalizations $f(0)=1$, $f(-1)=0$, $B^c$ and $B$ are the
non-universal magnetization amplitudes at the critical isotherm and at
the coexistence curve:
\begin{align}
\vec{M} &= B^c \vec{H} H^{1/\delta-1}, \qquad t=0,\\
M &= B (-t)^\beta, \qquad t<0, \ H\rightarrow 0.
\end{align}
The equation of state can also be written in the form:
\begin{equation}
\label{scalingF}
\vec{H} = ab \frac{\vec{M}}{M} t^{\beta\delta} F(z), \qquad z \equiv b M t^{-\beta},
\end{equation}
where the non-universal constants $a$ and $b$ fix the normalization
on $F(z)$ such that for $z \rightarrow 0$
\begin{equation}
\label{expF}
F(z) = z + \frac{z^3}{3!} + \sum_{n \geq 3}\frac{r_{2n}}{(2n-1)!}z^{2n-1}.
\end{equation}
Inverting (\ref{scaling}) the scaling equation of state can be written as:
\begin{equation}
\label{scalingE}
\vec{M} = B_c \vec{H} H^{1/\delta-1} E(y),
\end{equation}
where $y \equiv (B/B_c)^{1/\beta}tH^{-1/(\beta\delta)}$ is a scaling variable.
All the functions $f(x)$, $F(z)$ and $E(y)$ are universal.

\section{Approximate equation of state}
\label{approx}
The parametric representation
\begin{equation}
\label{par}
\left\{
\begin{aligned}
M &= m_0 R^\beta m(\theta)\\
t &= R (1-\theta^2)\\
H &= h_0 R^{\beta\delta}h(\theta)
\end{aligned}
\right.
\end{equation}
($m_0$ and $h_0$ are normalization constants)
implements the known analytical and scaling properties of the equation of state.
$R$ is a nonnegative variable which measures the distance from the
critical point. The functions $h(\theta)$ and $m(\theta)$ are odd and
are conventionally normalized so that $h(\theta)=\theta + O(\theta^3)$
and \mbox{$m(\theta)=\theta + O(\theta^3)$}, for $\theta \rightarrow 0$.
As can be seen from (\ref{par}), the line $\theta=0$ corresponds to
the high-temperature phase, while on the critical isotherm
$\theta=1$. The coexistence curve is given by $\theta=\theta_0$, where $\theta_0$ is the
first positive zero of $h(\theta)$.
Near $x=-1$ (coexistence curve), $f(x)\simeq c_f (x+1)^2$. This can be
satisfied if $h(\theta) \sim (\theta-\theta_0)^2$, for $\theta \rightarrow \theta_0$.

We start from the high-temperature, small magnetization, expansion
$z\rightarrow 0$ of $F(z)$, eq. (\ref{expF}).
Then, using the representation  (\ref{par}), we perform an analytic
continuation to the low-temperature phase.

We introduce two approximation schemes:
\begin{equation}
\begin{split}
&A: \left\{
\begin{aligned}
  m(\theta) &= \theta \left( 1+ \sum_{i=1}^n c_i\theta^{2i}\right)\\
  h(\theta) &= \theta \left( 1- \frac{\theta^2}{\theta_0^2} \right)^2
\end{aligned}
\right.\\
&B: \left\{
\begin{aligned}
  m(\theta) &= \theta \\
  h(\theta) &= \theta \left(1-\frac{\theta^2}{\theta_0^2}\right)^2\left( 1+ \sum_{i=1}^n c_i\theta^{2i}\right).
\end{aligned}
\right.
\end{split}
\end{equation}

For $n=0$ the two schemes are the same. In both cases, the
coefficients $c_i$ and $\theta_0$ are determined by imposing that the
equation of state, for $z \rightarrow 0$ or, equivalently,
$\theta \rightarrow 0$, reproduces the expansion (\ref{expF}).
We have considered both schemes for $n = 0$, $1$. We use field-theoretical estimates of $r_6$ and $r_8$, obtained by analyzing the perturbative series \cite{potential,ON}.
For the critical exponents we use the Monte Carlo estimates of \cite{exponentsO4} for $N = 4$
and field-theoretical estimates \cite{ON} for the other models.
Consistence of the whole computation requires the coefficients $c_i$ to be small.
Moreover, the Jacobian of (\ref{par}) must not vanish in the interval $[0, \theta_0]$.

\section{Results}
\begin{figure}
\includegraphics[width=0.88\linewidth, keepaspectratio]{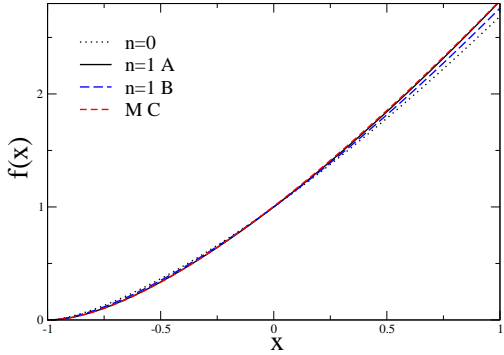}
\caption{The scaling function $f(x)$ for the $O(4)$ model. A comparison with the Monte Carlo result of ref.~\protect\cite{MCO4} is shown. From \protect\cite{O4}.}
\label{fx}
\end{figure}

\begin{figure}
\includegraphics[width=0.95\linewidth, keepaspectratio]{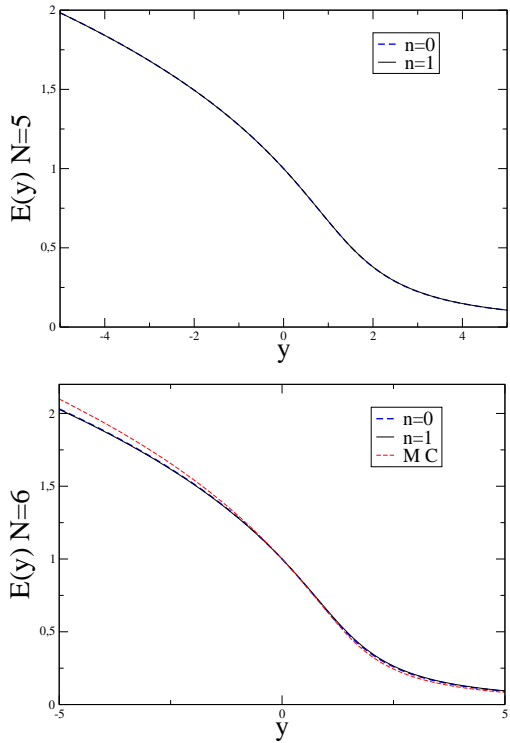}
\caption{The scaling function $E(y)$, for models $N=5$ (up) and $N=6$ (down). For the $O(6)$ model, we also show a comparison with the Monte Carlo result of ref.~\cite{MCO6}. From \protect\cite{ON}.}
\label{Ey}
\end{figure}

We show in figure \ref{fx}, taken from \cite{O4},
the scaling function $f(x)$ for the $O(4)$ model, as obtained with the
$n=0$, $n=1\ A$ and $n=1\ B$ schemes. We also show a comparison with the Monte Carlo result of ref.~\cite{MCO4}.
For $N>4$, scheme $B$ did not work, failing to satisfy consistency conditions (see end of section \ref{approx}).
In figure \ref{Ey}, taken from \cite{ON},
we show the scaling function $E(y)$ for the $O(5)$ and $O(6)$ models; $n=1$ refers to scheme $A$. For the $O(6)$ model, we also show a comparison with the Monte Carlo result of ref.~\cite{MCO6}.

From the approximate scaling equation of state,
it is possible to obtain several amplitude ratios.
We report in table \ref{ratios}
two of the most important ones: the ratio of the amplitudes of the specific heat $U_0$ and $R_\chi \equiv C^+ B^{\delta - 1}/(B^c)^\delta$, where $C^+$ is the amplitude of the susceptibility in the high-temperature phase \cite{review}.
We also show the same quantities for the $N=32$, $64$ models, along with a comparison with the large-$N$ expansion.
In the $N \rightarrow \infty$ limit, the $n=0$ scheme becomes exact \cite{largeNeq};
for the $O(32)$ and $O(64)$ models we find a good agreement with the large-$N$ results.

\begin{table}
\caption{Universal amplitude ratios.}
\begin{tabular}{lll}
\hline
$N$ & $U_0$ & $R_\chi$ \\
\hline
4   & 1.91(10)  & 1.12(11)\\
5   & 2.2(2)    & 1.2(1)\\
6   & 2.5(2)    & 1.15(9)\\
32  & 1.5(5)  & 0.94(1) \\
    & $1.47^*$  & $0.940^*$ \\
64  & 1.5(4) & 0.968(4) \\
    & $1.47^*$  & $0.9701^*$\\
\hline
\end{tabular}\\
\label{ratios}
$^*$Large-$N$
\end{table}

In conclusion, we note that the results obtained in the $n=1$ schemes are close to those in the $n=0$ scheme, thus supporting the effectiveness of the approximation scheme. For a more detailed discussion, as well as for other results, see \mbox{refs.~\cite{O4,ON}}.

\end{document}